\documentclass[]{article}
\RequirePackage{filecontents}
\usepackage{graphicx}
\usepackage{url}
\usepackage[T1]{fontenc} 
\usepackage[utf8]{inputenc} 
\usepackage[english]{babel} 
\usepackage[sort,comma,super,authoryear]{natbib}
\usepackage{amsthm}
\usepackage[hidelinks]{hyperref}

\theoremstyle{remark}
\newtheorem{remark}{Remark}

\title{An electronic data management and analysis application for ABET accreditation}
\author{Adeel Sabir\footnote{Corresponding author. Email: adeel\textunderscore sabir@hotmail.com}, Nisar A. Abbasi, Md Nurul Islam\\{\em Department of Electrical Engineering}\\ {\em University of Hafr Al Batin, Hafr Al Batin, Saudi Arabia}}
\date{}
\begin{document}
\maketitle
\begin{abstract}
This article presents an application developed for electronically managing and analyzing assessment data for ABET accreditation purposes using Microsoft Access. It facilitates the data entry, analysis and record-keeping for criterion 3 and 4 of the ABET engineering accreditation guidelines, which are arguably the most important, documentation-intensive and complex requirements in the entire process. Users can systematically manage large amounts of assessment data, conveniently run various queries and reports using pre-specified filters, and use them in analyzing the strengths, weaknesses and critical areas of the educational program. For closing the assessment cycle loop, the electronic tool also provides the ability to manage verbal feedback and observations for planning remedial actions and continually improving the program. The use of the application is demonstrated through illustrative examples on data gathered over multiple academic terms. The program and its user guide are available to educators and evaluators\footnote{\href{https://www.dropbox.com/s/tpeftggj06zsjpn/ABET_App_distribution.zip?dl=0}{Download link}}. \newline\newline
\end{abstract}

\section{Introduction} \label{sec:introduction}
The Accreditation Board of Engineering and Technology (ABET) is an internationally renowned body for evaluating engineering programs and many institutions all over the world seek its authentication to highlight the quality of their engineering education. However, preparing for ABET accreditation is an extensive exercise marked by the collection and maintenance of large amounts of assessment data and documentation over extended periods. Not only must the data be properly organized, but also objectively analyzed to extract meaningful information relevant to the academic program and the enrolled students. Criterion 3 and 4 are perhaps the most overwhelming yet important sections in the ABET Criteria for Accrediting Engineering Programs (CAEP) document \citep{abet_eac2017} as they relate directly to the curriculum \citep{Miller2016}. They are also the ones requiring the bulk of documentation since they deal with student outcomes and documented improvement processes. Maintaining and analyzing the paperwork associated with these criterion is a nontrivial task and can be quite laborious and cumbersome. While the workload and efforts related to ABET assessment can be overseen by a subcommittee of faculty members and administrators, all faculty members must have a working knowledge of the accreditation process and be able to intelligently talk about it on the day of the accreditation visit \citep{Wear2012}. It is only possible if they are actively engaged in the process and do not feel overburdened by the workload associated with carrying out ABET assessment tasks.   

Using electronic tools for data management and analysis can make the assessment process much more efficient and, in turn, encourage wider faculty participation. Additionally, substituting the hard copies documents with electronic ones also carries environmental benefits and saves large amounts of money in printing costs. A number of electronic applications have been presented in existing literature to automate the documentation-heavy areas ABET's outcome based assessment. Some of them are based on existing commercial packages \citep{Kerr2011,eLumen2018,Eltayeb2012} while a majority consists of tools developed in-house \citep{Petrova2006,Essa2010,Trytten2010,Zahorian2011,Shankar2013,Ibrahim2015,Ismail2016}. In \citep{Petrova2006}, a web-based system is presented to store the comprehensive assessment data (e.g., student samples, surveys, course samples, courses-to-outcome maps, etc.) and present it conveniently to the accreditation evaluators, allowing them to navigate through it in a handy manner. In \citep{Essa2010}, a web-based tool called ACAT is developed enabling the faculty to generate course assessment reports for ABET accreditation. An electronic data analysis tool is proposed in \citep{Trytten2010} to efficiently collect, manage and display the instruments used in the assessment i.e., course syllabuses and policies, assignments and samples of student work. Another web database application is presented in \citep{Zahorian2011} built on an existing paper-based method, allowing the faculty to enter assessment data through web-based forms into a database and numerically analyze its statistics in an automated way. In \citep{Kerr2011}, the adoption of web-based system called eLumen \citep{eLumen2018} is outlined to obtain accreditation from the Canadian Engineering Accreditation Board (CEAB) through outcomes-based assessment. A network-based data management and assessment tool is created in \citep{Kelly2012} using Microsoft Access to document and evaluate course syllabi, lecture materials and samples of student work in electronic format. Similarly, several recent publications have highlighted the importance using electronic means to make cumbersome tasks involved in ABET assessment more efficient and effective (see, for example \citep{Eltayeb2012,Shankar2013,Ibrahim2015,Ismail2016}). 

There are advantages and disadvantages to choosing between a commercially available tool or developing a custom application. While ready-made packages like eLumen \citep{eLumen2018} or EvalTools \citep{evaltools2018} save development time and efforts, the associated costs and faculty training may prevent an academic department from using them \citep{Petrova2006}. On the other hand, building a custom tool from scratch offers the ability to tailor it to the program's needs and integrate it into the assessment processes and methodologies that are already in place - benefits that may not be reaped by the former approach. The downside to custom-building an application is the need for software engineering skills, choice of proper development tools, and requirements of personnel, resources and time. In fact, a majority of custom-built tools mentioned above are built through collaboration between academics and professionals that appear to have computer science backgrounds and the tools availed in application development require proficiency in software skills  \citep{Petrova2006,Essa2010,Zahorian2011,Shankar2013,Ibrahim2015}. For academics from other disciplines having little to no knowledge of software development, such requirements may prove to be limiting factors in migrating to electronic platforms for their assessment needs. Therefore, there is a critical need put forth a solution that can be built and customized by academics and professionals with less stringent software skill requirements.

Microsoft Access is a database management system (DBMS) that combines relational database management with a graphical user interface (GUI) and software-development tools \citep{accesswiki2018}. It is designed for ease of use and includes many powerful features offered by advanced DBMSs like SQL Server and Oracle. Since it is a part of the Microsoft Office Professional Suite, it is readily available in most organizations and universities as a licensed software. Having a user-friendly interface similar to the commonly used Microsoft Excel, plentiful availability of self-paced online learning resources \citep{accesstutorial_tutorialspoint_2018,accesstutorial_microsoft_2018}, and possessing many powerful capabilities offered by high-end DBMSs, Microsoft Access is, in the authors' opinion, a highly appropriate tool that can readily bridge the gap between an average user and rapid development of a customized relational database application. It has been successfully employed in various academic projects like expressing solution concentrations \citep{Serban2010} and plagiarism detection \citep{McCart2008}, as well as ABET accreditation and record keeping \citep{Cliver2011,Kelly2012,Scales1998}. Reference \citep{Cliver2011} should be of particular interest to the readers where the authors have highlighted some technical aspects of Microsoft Access like tabular relationships, data entry forms, and report generation for fulfilling ABET requirements.

While the authors in \citep{Cliver2011,Kelly2012} have effectively demonstrated the use of Access in automating ABET assessment data management, the tools presented therein do not appear to offer adequate capabilities that can be employed to cover all aspects of the assessment process through a single, unified user interface. On the other hand, solutions that offer such capabilities e.g., \citep{Zahorian2011,Ibrahim2015,Shankar2013}, are designed using professional development tools. Hence, there is still ample room to demonstrate that a powerful integrated application can be developed using a readily available, GUI-based application that can handle the bulk of the demanding tasks in an outcomes-based assessment activity. The key contribution of this article is that a comprehensive, multi-user application to efficiently maintain and analyze ABET accreditation data, built entirely in Microsoft Access, is presented. It allows efficient data management for the documentation-intensive criterion 3 and 4, and enables the users to manage and view the assessment data in the form of customized reports for the purposes of qualitative and quantitative analysis. The activities associated with these criterion are automated and systematized to a much greater extent than some existing Access-based proposals on ABET assessment \citep{Cliver2011,Kelly2012,Scales1998}. The application is currently being used by the Department of Electrical Engineering (EE) at the authors' institution. A barebone version of the application and its user-guide can be downloaded here\footnote{\href{https://www.dropbox.com/s/tpeftggj06zsjpn/ABET_App_distribution.zip?dl=0}{Download link}}. 

The remainder of this article is organized as follows: the methodology adopted for assessment is described in Section \ref{sec:method}. Details of the Access application are given in Section \ref{sec:application}. Results derived from using the application are reported and discussed in Section \ref{sec:examples}. Concluding remarks and future work are given in Section \ref{sec:conclusion}.  

\section{Methodology} \label{sec:method}
\subsection{In a nutshell}
\begin{figure}[t!]
	\centering
	\includegraphics[width=\linewidth]{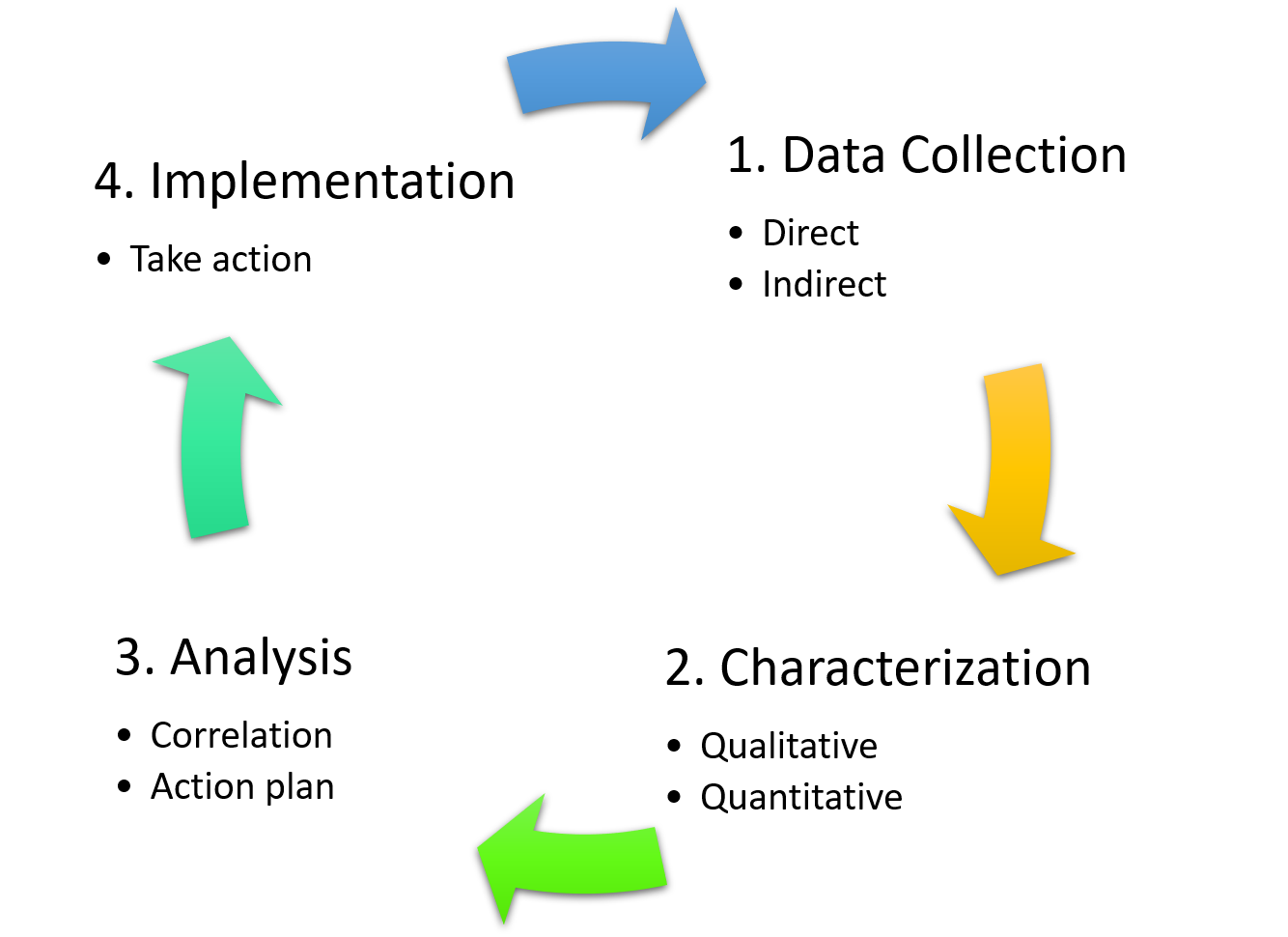}
	\caption{Assessment process in a nutshell.}
	\label{fig:nutshell}
\end{figure}

The methodology of the assessment process can be summarized in four main steps depicted graphically in Figure \ref{fig:nutshell}. 
First, the assessment data is collected through direct and indirect means from the faculty members, enrolled and graduating students, and the alumni. It is then characterized into quantitative and qualitative data. Next, the gathered information is analyzed to identify correlations, eminent trends, and areas needing attention and a corrective action plan is devised. This step is succeeded by implementation, whereby the necessary actions identified in the analysis phase are implemented. These steps are repeated in the next iteration. There are two terms, or semesters, in an academic year excluding the optional summer term. A complete assessment cycle comprises of four terms, or two academic years. Hence, the activities shown in Figure \ref{fig:nutshell} are carried out four times in an assessments cycle. Further details about these activities are given in the following subsections.

\subsection{Data collection}
Primary sources of assessment information include faculty members, enrolled and graduating students, and alumni. The collected information is categorized into direct and indirect data types based on its source. The method is similar to the ones reported in \citep{Alyahya2012,Al-Yahya2013,Abu-Jdayil2010}. 
\subsubsection{Direct data}
Direct data comes from the faculty members' evaluation of student performance in their respective courses, and their verbal feedback, observations and recommendations on various aspects of a course or the program. 
\subsubsection{Indirect data}
Indirect assessment data is gathered primarily through surveys from the enrolled students, graduating students and the alumni. Survey participants are asked a series of multiple choice and subjective questions regarding their perception on their degree of accomplishment of course outcomes (COs) and student outcomes (SOs). They also provide their general observations and feedback on different course aspects or the program as a whole. Enrolled and graduated students are survey each term while the alumni surveys are periodically conducted over longer intervals.

\subsection{Characterization} \label{ssec:characterization}
Characterization includes classifying the assessment information into quantitative and qualitative types. 
\subsubsection{Quantitative data}
\begin{figure}[t!]
	\centering
	\includegraphics[width=\linewidth]{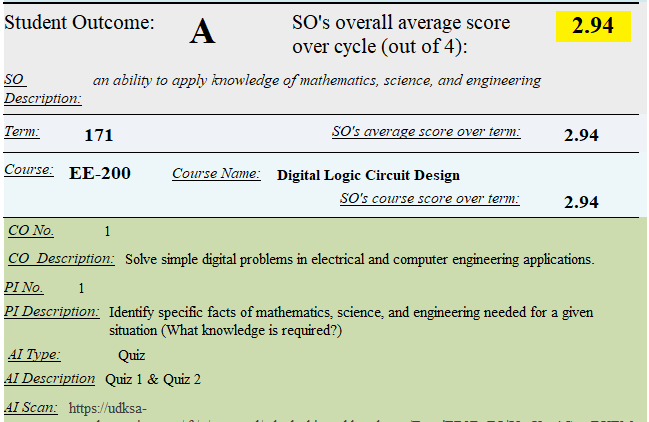}
	\caption{SO A to CO 1 mapping for EE-200: Applied Electromagnetics course for direct assessment.}
	\label{fig:map2}
\end{figure}

\begin{figure}[t!]
	\centering
	\includegraphics[width=\linewidth,trim=3cm 2cm 18cm 4.75cm,clip]{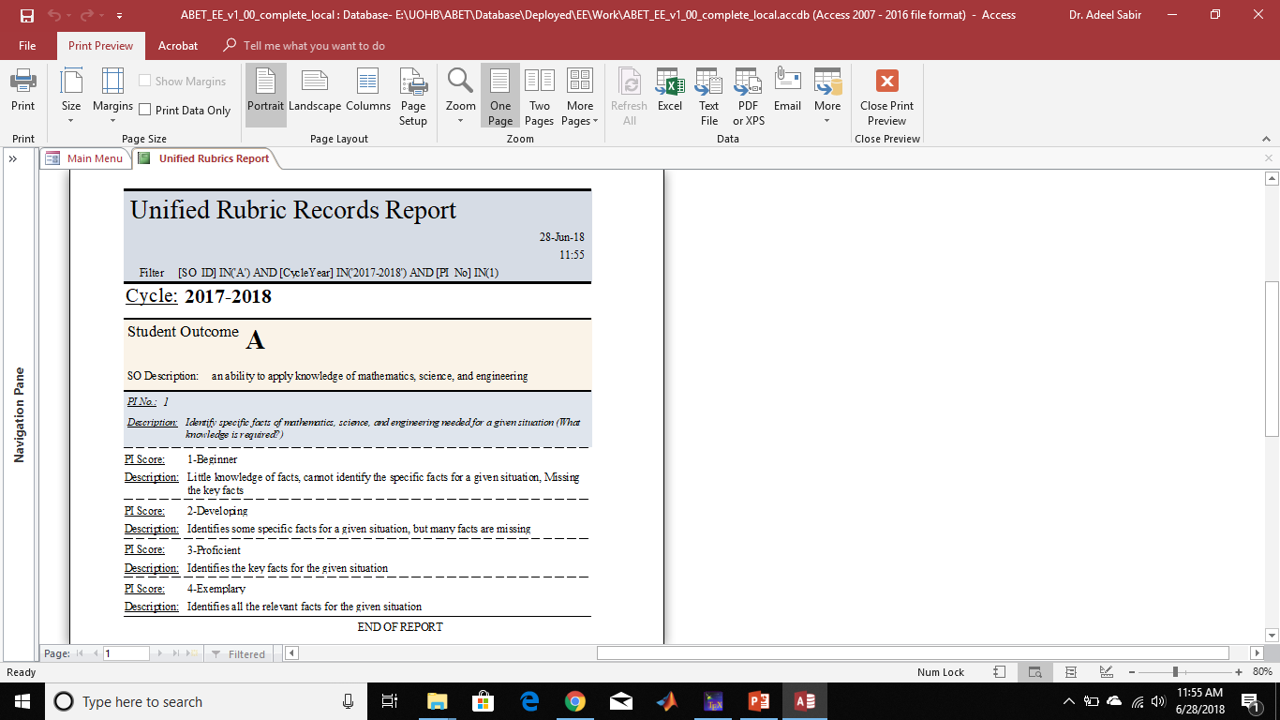}
	\caption{Description of a PI and its proficiency levels for SO A.}
	\label{fig:rubricdefA1}
\end{figure}

Quantification of data involves its evaluation and mapping to a numerical score. For direct assessment, COs are mapped to ABET SOs A through K \citep{abet_eac2017}. A course instructor evaluates a sample group of students on these mapped SOs using performance indicators (PIs) - subjective descriptions of the abilities that measure performance on a specific outcome, and student scores on the appropriate assessment instruments (AIs) - the means used to assess a student's performance in an outcome. The scores obtained by each student in the group are normalized on a scale of 1-4 corresponding to the proficiency levels, and averaged to obtain a numerical value for each SO mapped to a CO. A threshold of $60\%$ is set as the minimum acceptable score, below which corrective actions are identified for its improvement in future assessment iterations. A sample of direct assessment showing the mapping and score for the course EE-200: Digital Logic Circuit Design, is shown in Figure \ref{fig:map2}. Here, SO A is mapped to CO 1 and is assessed using the AIs of quiz 1 and quiz 2. The SO gets an averaged numerical score of 2.94 out of 4 for the assessed student sample in that term. Detailed description of a PI proficiency levels and scores for SO A are shown in Figure \ref{fig:rubricdefA1}.  

Indirect assessment data is quantified using the responses to the multiple-choice survey questions formulated directly in terms of the accomplishment of SOs and COs. Responses to these questions are recorded as \textit{strongly agree}, \textit{agree}, \textit{neutral}, \textit{disagree} and \textit{strongly disagree}, with the choice \textit{strongly agree} indicating that an outcome is met to a high degree. These responses are mapped to a scale of 1-5 where 5 represents the highest degree of achievement of a SO. A weighted average of scores is calculated to get an indirect numerical score for each SO. Similar to the direct assessment, a minimum threshold of $60\%$ is set as the acceptable. A score below the threshold prompts identification and implementation of corrective actions. 

The numerical scores thus calculated from direct and indirect assessments constitute the quantitative data.

\subsubsection{Qualitative data}
Verbal feedback and recommendations from the faculty members, students and alumni, based on their observations and experiences during teaching, enrollment and post-graduation, are systematically recorded and characterized as the qualitative assessment data. 

\subsection{Analysis, correlation and action plan}
Following the characterization phase, the accumulated information is analyzed by the department's ABET committee to identify correlating patterns and prominent trends. Numerical SO scores from direct and indirect assessments are compared, and the qualitative data and feedback are subjectively examined. A rigorous analysis serves to find links between faculty and student perceptions on SO accomplishment, uncover key areas that require attention, and helps in formulating a comprehensive agenda for continually improving the program. Active measures derived from data analysis are categorized into immediate and long-term actions to be taken. Status of pending tasks from previous assessment iterations are also checked and a thorough action plan is devised. 

\subsection{Implementation}
\begin{figure}[t!]
	\centering
	\includegraphics[width=\linewidth]{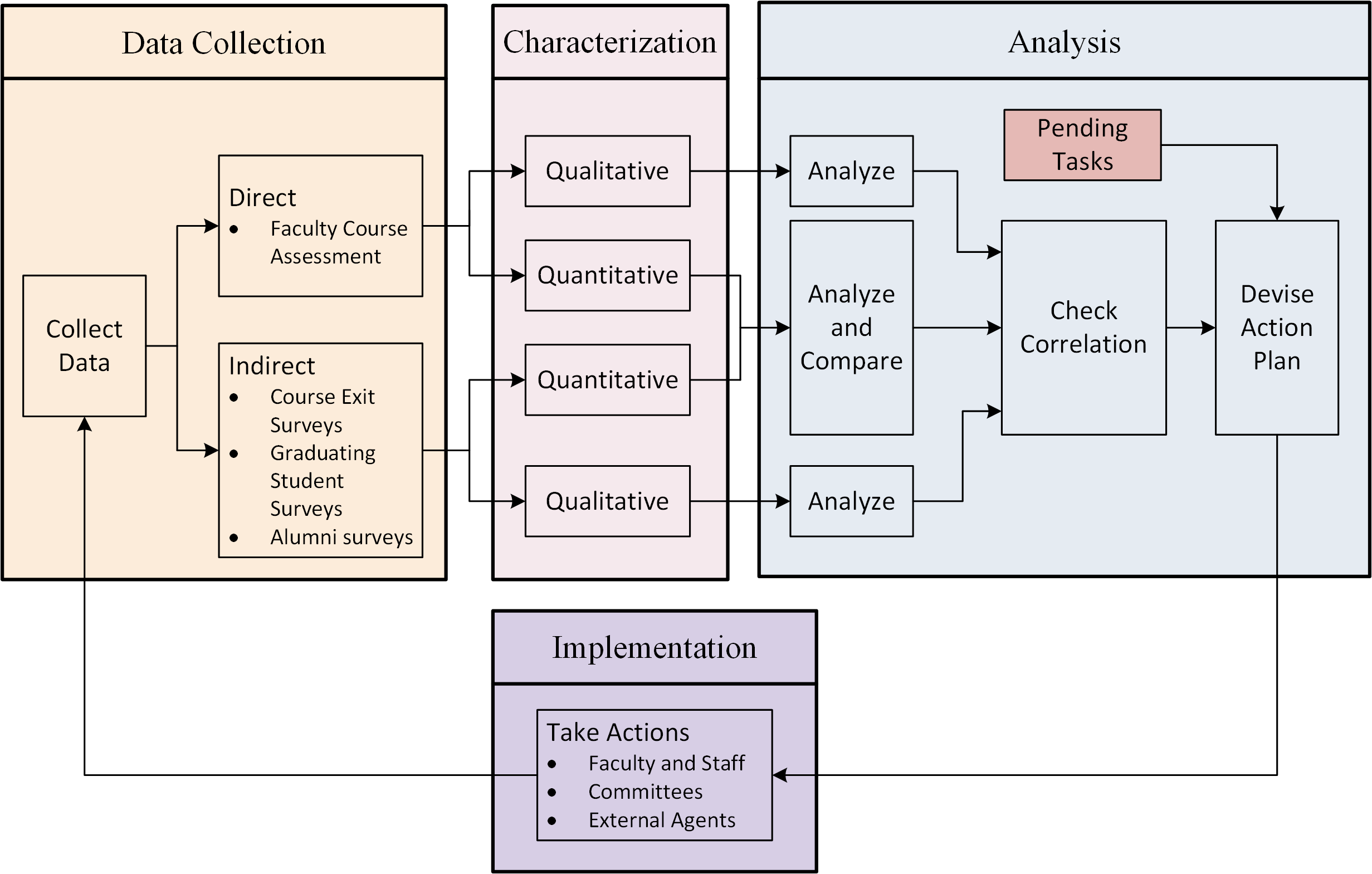}
	\caption{Workflow of the assessment process.}
	\label{fig:assessment2}
\end{figure}

The devised action plan is implemented through individual and collaborative actions and initiatives involving various entities internal and external to the university. Internal entities include individuals such as faculty members, lab instructors and technical staff, or groups of individuals such as special purpose committees e.g., the ABET accreditation committee or the program revision committee. Inter-departmental committees, university management or the internal board of directors may also be involved as necessary. Examples of external entities include the industrial partners where students undertake co-op training, or equipment vendors and trainers that provide technical support and training to the faculty and students on the tools used in education. Workflow of the assessment process is illustrated in Figure \ref{fig:assessment2}.

\section{ACT: The ABET accreditation tool} \label{sec:application}
\subsection{Tool architecture}
\begin{figure}[t!]
	\centering
	\includegraphics[width=\linewidth]{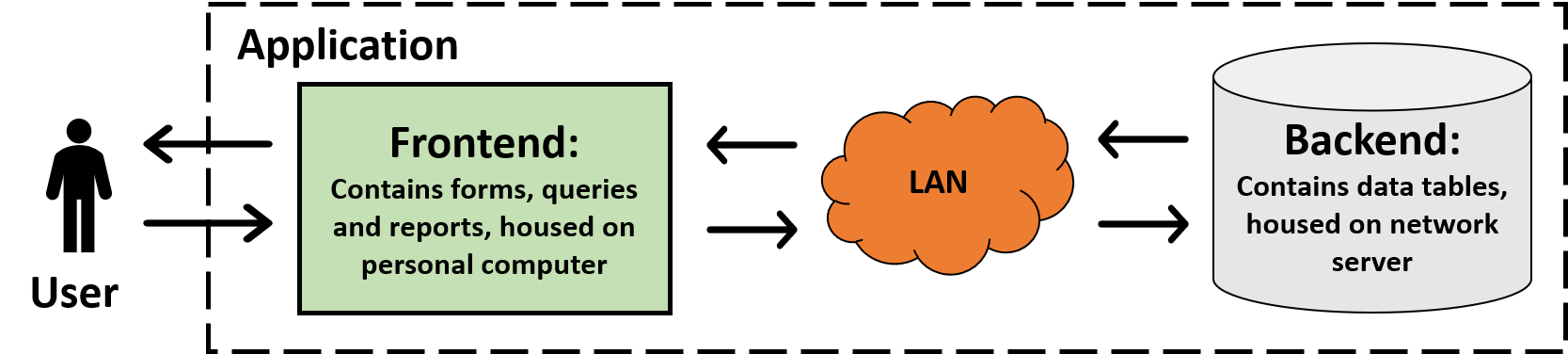}
	\caption{Tool architecture.}
	\label{fig:app}
\end{figure}

The assessment data is maintained in a database application, named as the Accreditation Tool (ACT), designed in Microsoft Access. The users interact with it through a frontend that houses various interactive data management forms along with querying and report generating tools. The data is stored in tables housed in the backend stored on a secure network server with controlled permissions. A copy of the frontend is distributed to the faculty to be used from their personal computers for interacting with the backend data. The architecture of the application is shown in Figure \ref{fig:app}.

\subsection{Getting started}
\begin{figure}[t!]
	\centering
	\includegraphics[width=\linewidth,trim=8cm 8cm 8cm 0,clip]{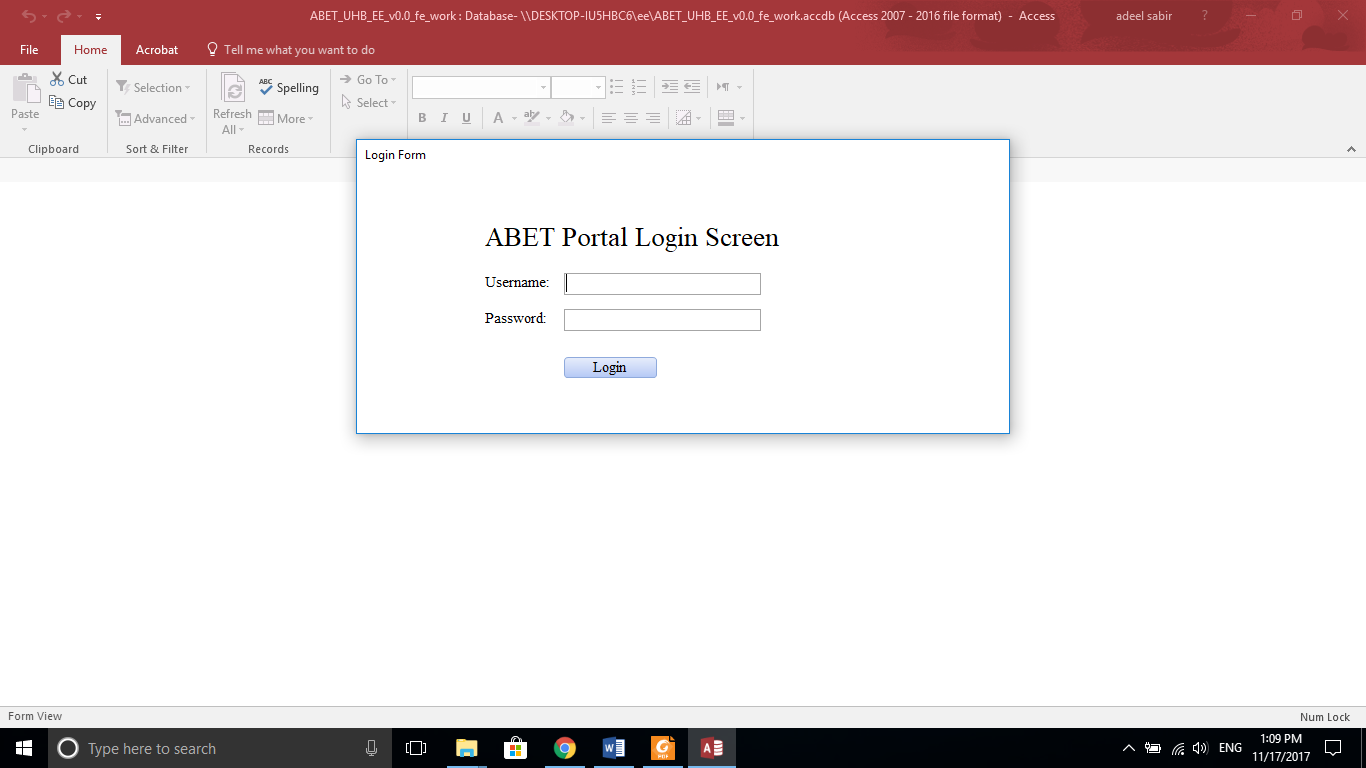}
	\caption{The login screen.}
	\label{fig:login}
\end{figure}

\begin{figure}[t!]
	\centering
	\includegraphics[width=\linewidth,trim=0 1cm 12cm 0,clip]{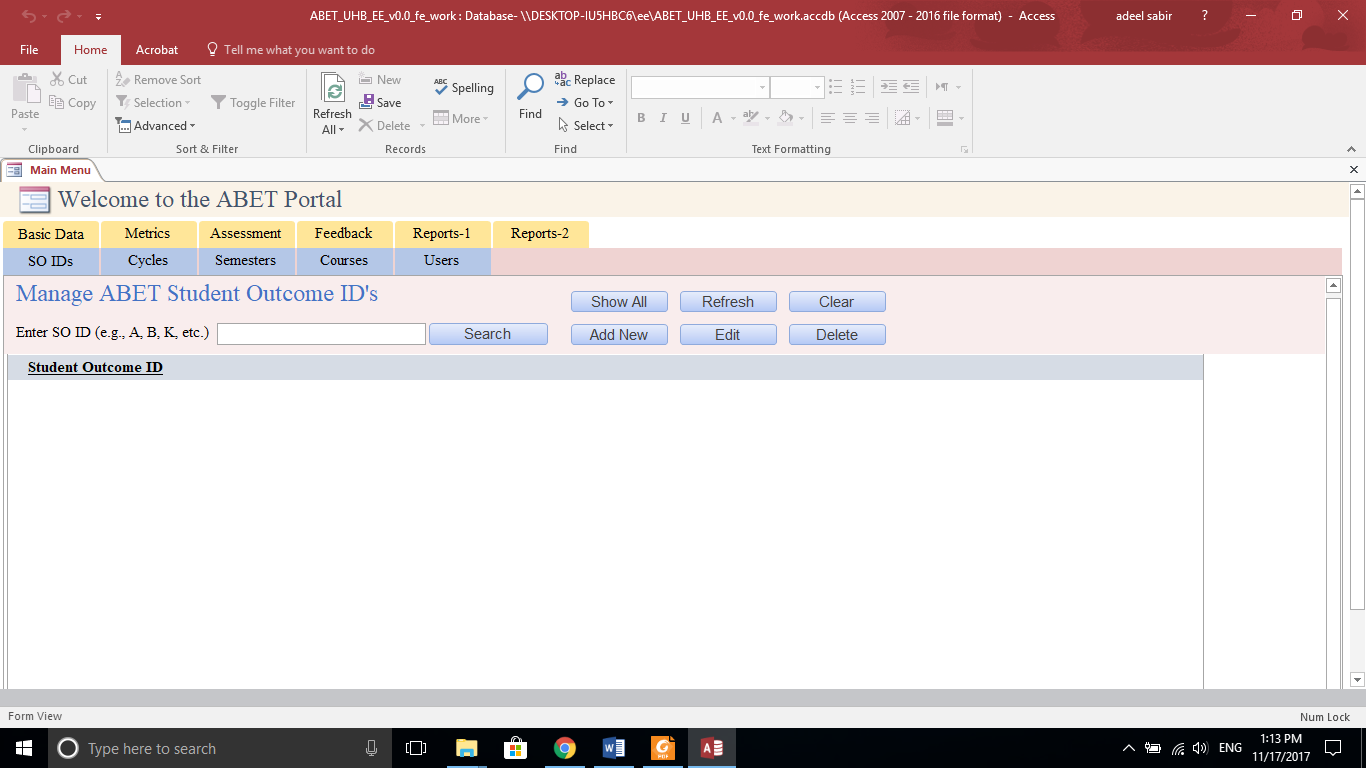}
	\caption{The main menu.}
	\label{fig:mainmenu}
\end{figure}

\begin{figure}[t!]
	\centering
	\includegraphics[width=\linewidth]{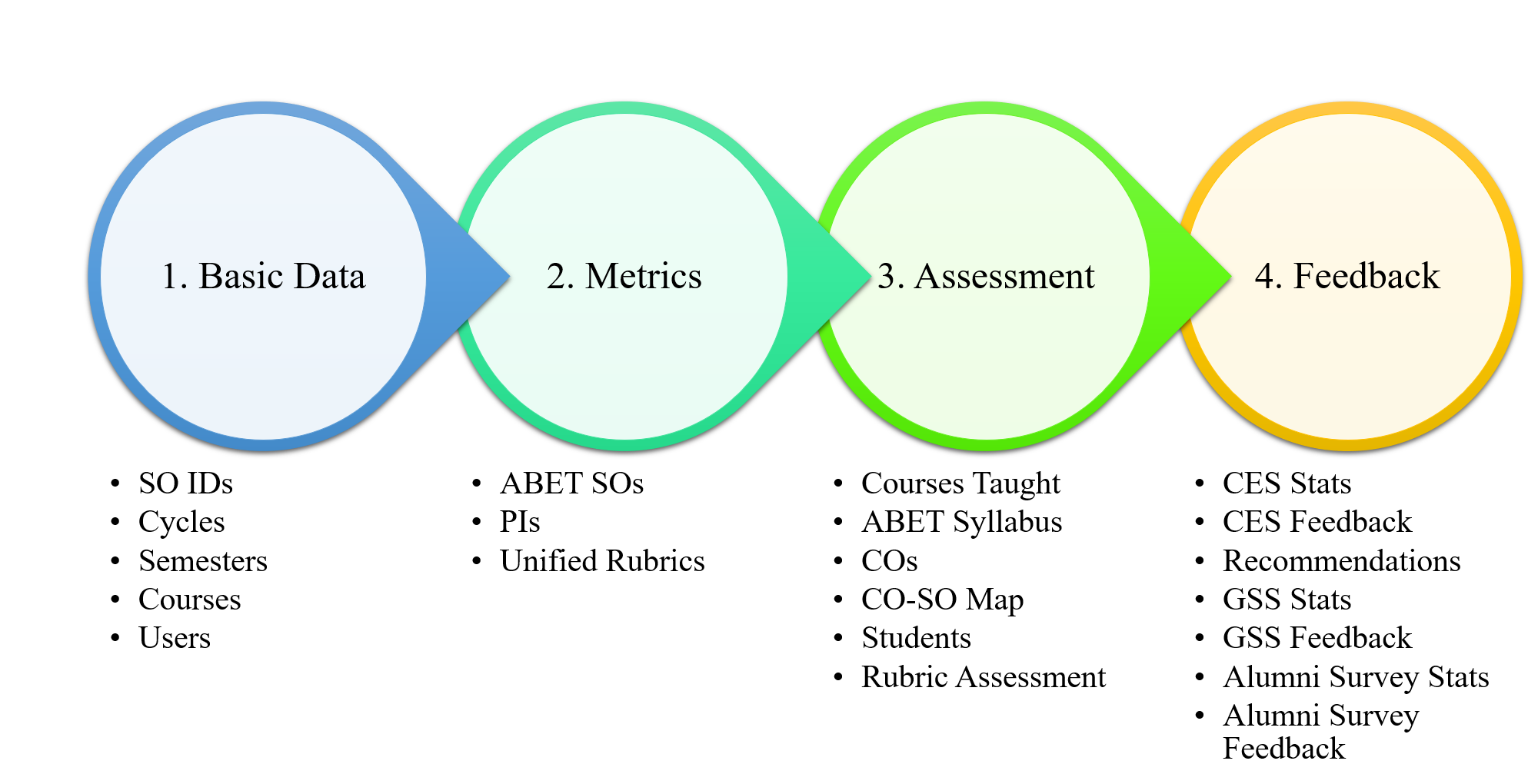}
	\caption{Workflow of the data-entry process.}
	\label{fig:dataentrysteps}
\end{figure}

Once ACT is opened through the frontend, the user is taken to a login screen and asked to enter a username and password. Upon successful login, the main screen appears containing six tabs. The login and main menu screens are shown in Figures \ref{fig:login} and \ref{fig:mainmenu}, respectively. Users can interact with the tool via interactive tabs labeled \textit{Basic Data}, \textit{Metrics}, \textit{Assessment}, \textit{Feedback}, \textit{Reports-1} and \textit{Reports-2}. They contain sub-tabs linked to various forms that handle data in the backend tables. Access allows users the option to place tabbed controls on forms through a graphical user interface (GUI), giving the application a user-friendly appearance. The tabs are arranged in the order the users are required to enter assessment data into the database.
Data-entry is completed in four sequential steps that include a number of sub-steps, each requiring specific inputs from the users. Workflow of this process is illustrated in Figure \ref{fig:dataentrysteps}.

\subsection{Basic data}
\begin{figure}[t!]
	\centering
	\includegraphics[width=\linewidth,trim=0 1cm 18cm 4.25cm,clip]{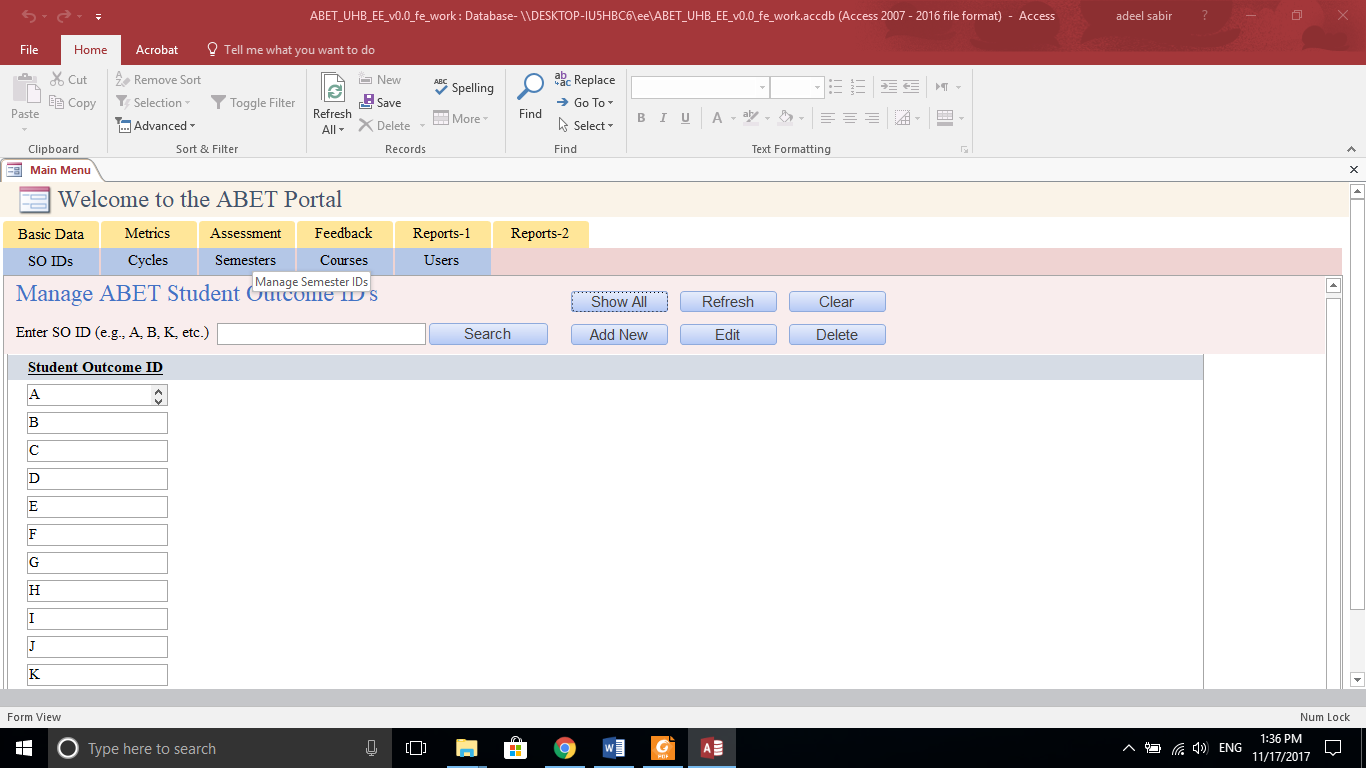}
	\caption{\textit{SO IDs} form showing existing records.}
	\label{fig:soid1}
\end{figure}

\begin{figure}[t!]
	\centering
	\includegraphics[width=\linewidth,trim=8cm 8cm 8cm 4.5cm,clip]{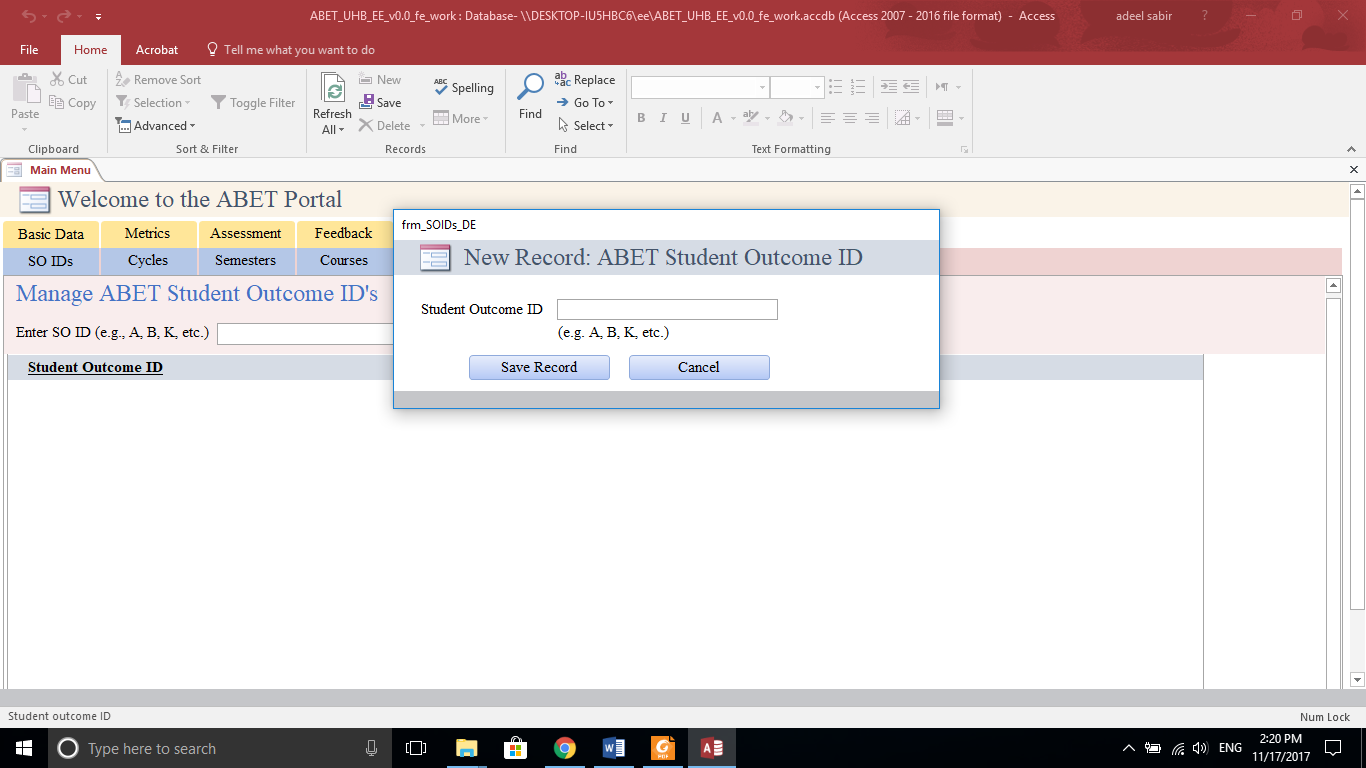}
	\caption{Adding a new record using the \textit{SO IDs} form.}
	\label{fig:soid2}
\end{figure}

\textit{Basic Data} is the first tab that opens following a successful login. It contains five different forms under the sub-tabs labeled \textit{SO IDs}, \textit{Cycles}, \textit{Semesters}, \textit{Courses} and \textit{Users}. Each of these forms allows the users to manage (view, add, edit, delete or search) specific assessment records. The functions available under the \textit{Basic Data} tab are designed to prepare the database for first usage and the associated data needs only be entered once per assessment cycle. The functions in these forms are listed below:
\begin{itemize}
	\item \textit{SO IDs}: manage the alphabetical student outcome labels (A-K)
	\item \textit{Cycles}: manage assessment cycle labels
	\item \textit{Semesters}: manage semester labels
	\item \textit{Courses}: manage course labels
	\item \textit{Users}: manage database users and their access levels
\end{itemize} 
Semesters and cycles are denoted by brief labels. For instance, the first term in an academic year 2017-2018 is denoted by the label `171'. An assessment cycle that began in term 171 will end in 182 and is denoted by the label `2017-2018', with the processes shown in Figure \ref{fig:nutshell} repeatedly executed in terms 171, 172, 181 and 182. Snapshots of the \textit{SO IDs} form with existing records and during adding a new record are shown in Figures \ref{fig:soid1} and \ref{fig:soid2}, respectively.        

\begin{remark}
At the time of writing this article, the authors' institution was following ABET accreditation criteria document 2017-2018 \citep{abet_eac2017}. Starting from 2019-2020 accreditation review cycle, \textit{Criterion 3 - Student Outcomes} will change \citep{abet_eac2018} - SOs A through K will be replaced by new aggregated outcomes 1-7. The application can readily be adopted to the new criteria once the COs have been mapped to the new ABET SOs, by merely using different SO labels e.g., 1-7 instead of A-K.
\end{remark}

\subsection{Metrics}
\begin{figure}[t!]
	\centering
	\includegraphics[width=\linewidth,trim=0cm 1cm 12cm 4.25cm,clip]{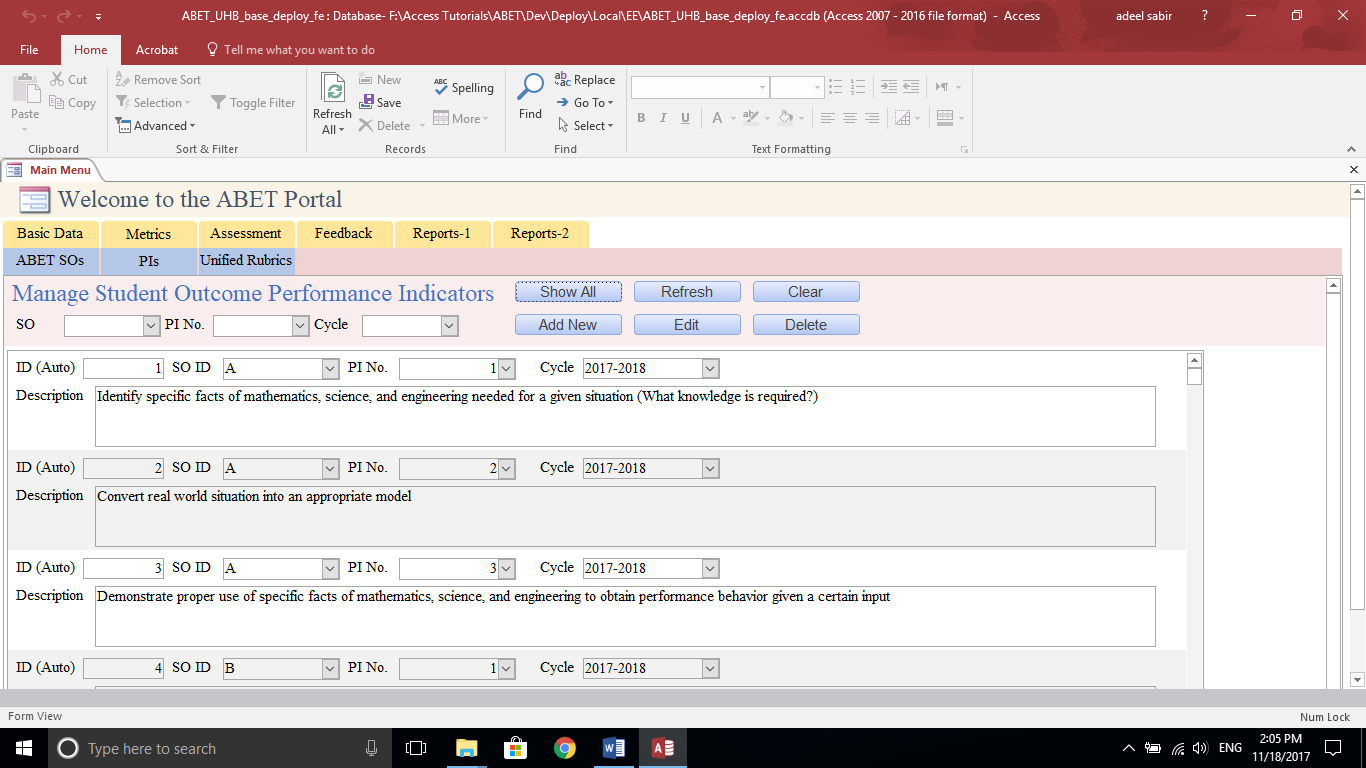}
	\caption{\textit{PIs} form showing existing records.}
	\label{fig:PI1}
\end{figure}

\begin{figure}[t!]
	\centering
	\includegraphics[width=\linewidth,trim=0cm 1cm 12cm 4.25cm,clip]{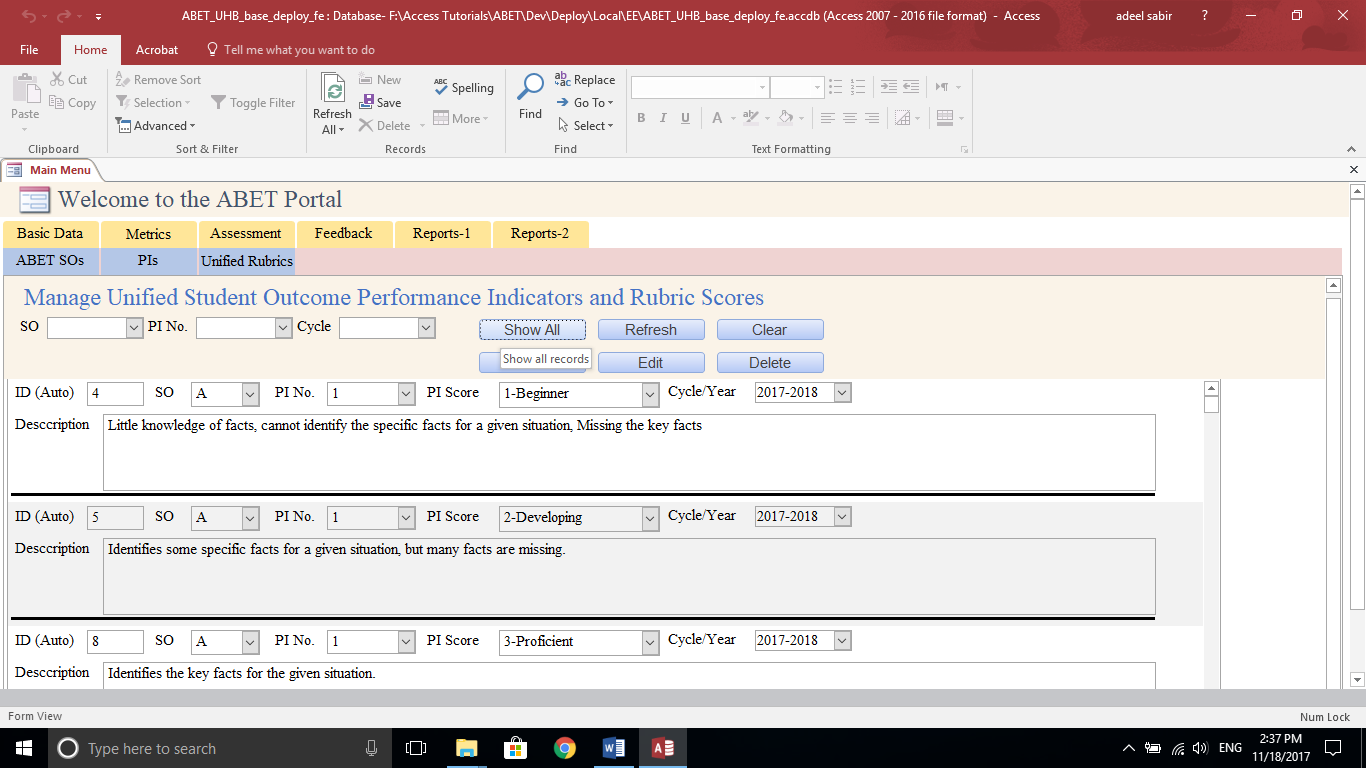}
	\caption{\textit{Unified Rubrics} form showing existing records.}
	\label{fig:rubrics1}
\end{figure}

\textit{Metrics} is the second main tab in ACT and contains three different forms labeled \textit{ABET SOs}, \textit{PIs} and \textit{Unified Rubrics}. They are designed to administer verbal descriptions of the metrics used in assessment. The form functionalities are described below:
\begin{itemize}
	\item \textit{ABET SOs}: manage verbal descriptions of ABET SOs
	\item \textit{PIs}: manage verbal descriptions of performance indicators for the ABET SOs
	\item \textit{Unified Rubrics}: manage verbal descriptions of performance rubrics for ABET SOs and PIs
\end{itemize} 
Figures \ref{fig:PI1} and \ref{fig:rubrics1} show the screenshots of the PI and rubric forms with existing records, respectively.

\subsection{Assessment} \label{ssec:assessment}
\begin{figure}[t!]
	\centering
	\includegraphics[width=\linewidth,trim=9.5cm 4cm 9.5cm 1cm,clip]{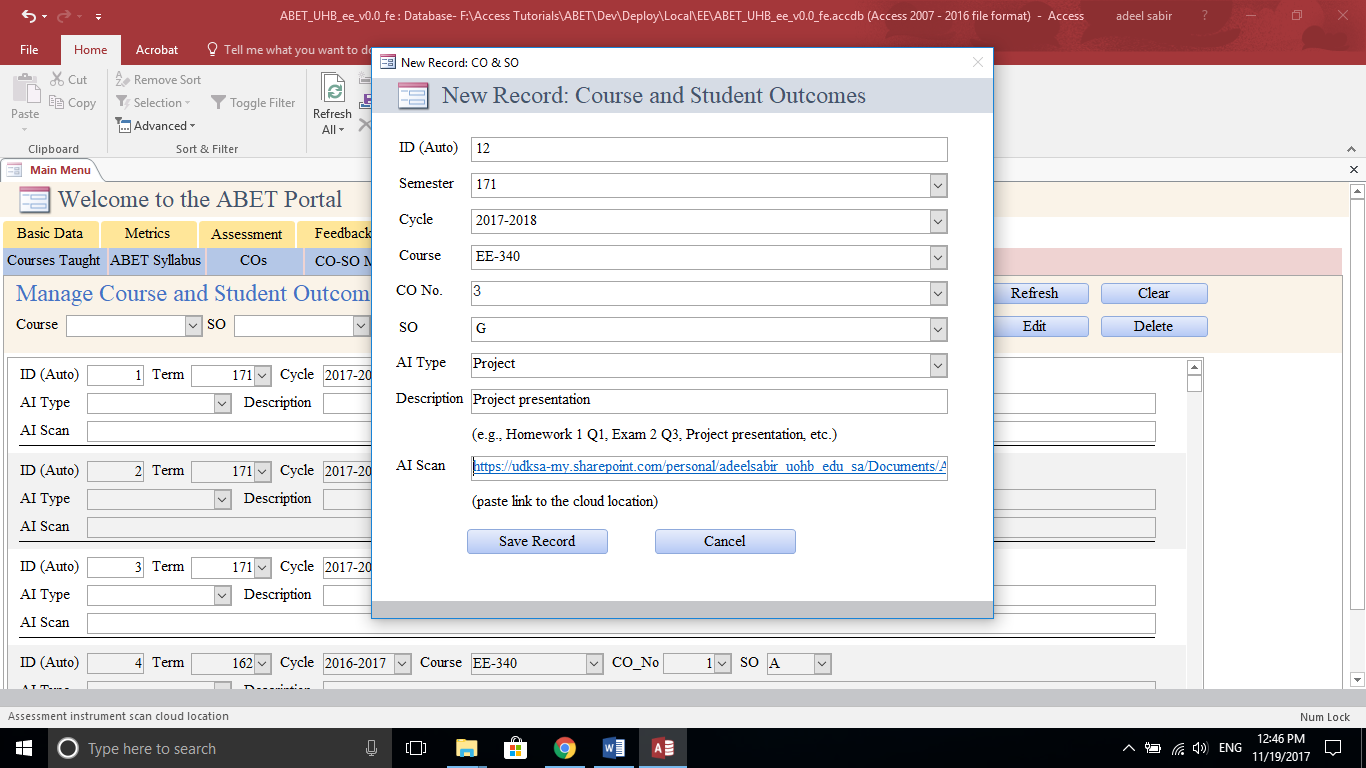}
	\caption{An example record entry in \textit{CO-SO Map} form.}
	\label{fig:COSO1}
\end{figure}

\begin{figure}[t!]
	\centering
	\includegraphics[width=\linewidth,trim=0cm 1cm 5cm 4.25cm,clip]{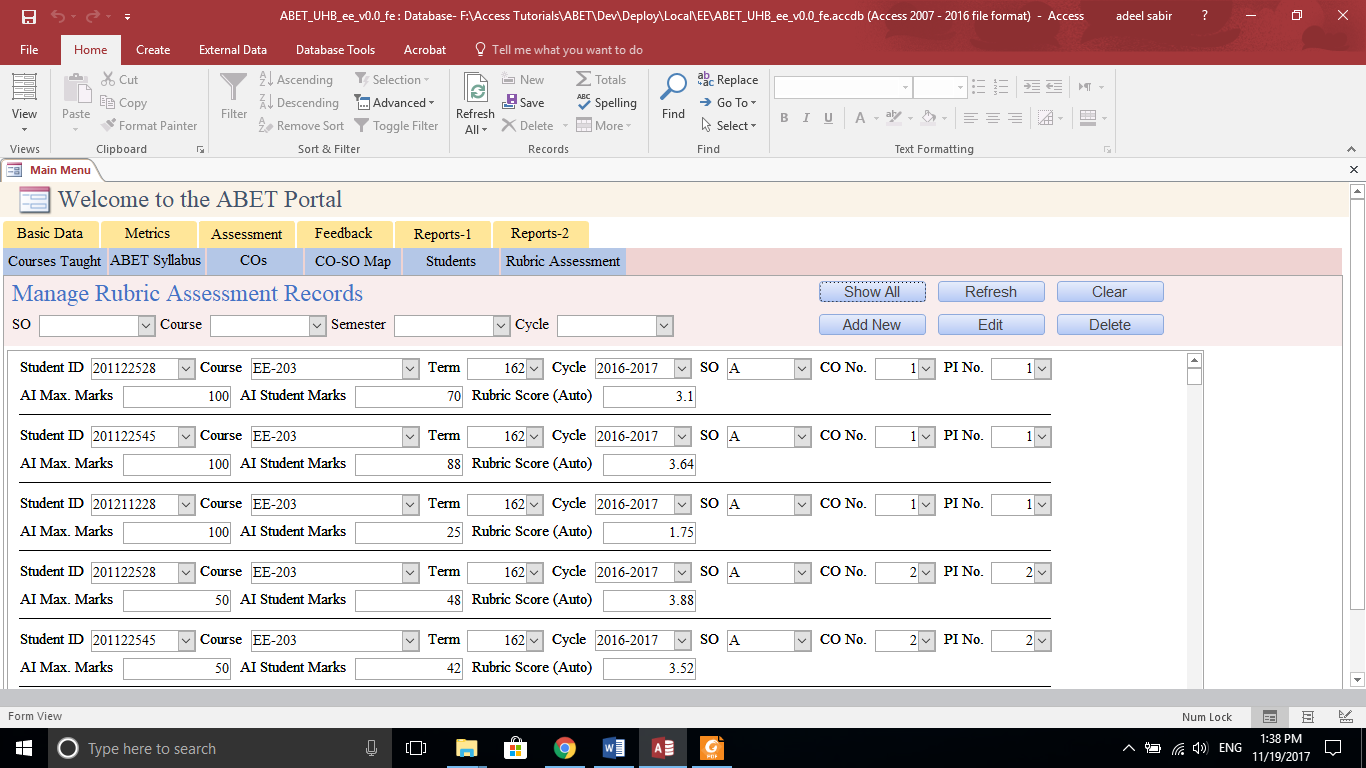}
	\caption{\textit{Rubric Assessment} form showing some existing records.}
	\label{fig:rubricassess1}
\end{figure}

\textit{Assessment} is the third main tab containing six different forms under the sub-tabs labeled \textit{Courses Taught}, \textit{ABET Syllabus}, \textit{COs}, \textit{CO-SO Map}, \textit{Students} and \textit{Rubric Assessment}. Functions available under the \textit{Assessment} tab are designed to manage records central to the direct assessment method. The forms provide the following functions:
\begin{itemize}
	\item \textit{Courses Taught}: manage courses taught by faculty
	\item \textit{ABET Syllabus}: manage ABET syllabi for offered courses
	\item \textit{COs}: manage course outcomes for offered courses
	\item \textit{CO-SO Map}: manage mapping between COs and SOs for offered courses
	\item \textit{Students}: manage student data e.g., IDs, names and expected graduation terms
	\item \textit{Rubric Assessment}: manage rubric assessments
\end{itemize} 
When mapping a CO to a SO, the \textit{CO-SO Map} form also allows the user to link a CO-SO relationship to a display material (scanned copies of student assignments, reports, exams, etc.,) to be presented to the evaluator. A scanned display document can be interactively linked to this mapping using the \textit{AI Scan} field. Moreover, while entering direct assessment data using the \textit{Rubric Assessment} form, the rubric score for each student is automatically calculated on a scale of 1-4 using an inbuilt equation. Figure \ref{fig:COSO1} displays the data entry form while a CO is mapped to a SO and Figure \ref{fig:rubricassess1} shows some rubric assessment records. 

\subsection{Feedback} \label{ssec:feedback}
\begin{figure}[t!]
	\centering
	\includegraphics[width=\linewidth]{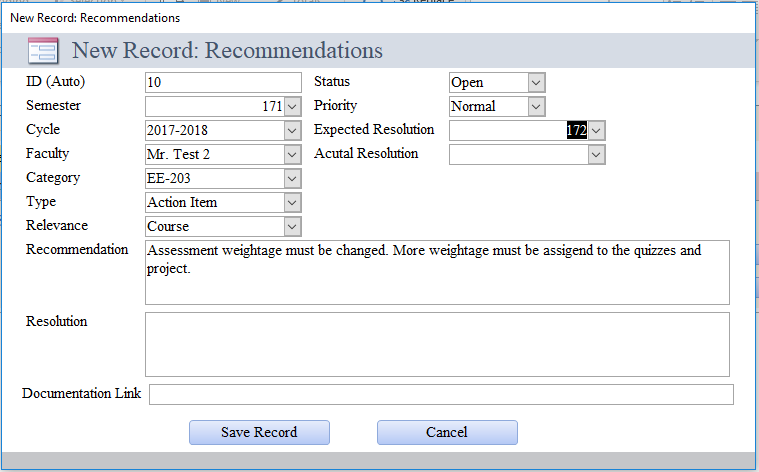}
	\caption{A sample record entry using the \textit{Recommendations} form.}
	\label{fig:recommend1}
\end{figure}

\textit{Feedback} is the fourth main tab in ACT. It contains seven different forms under the sub-tabs labeled \textit{CES Stats}, \textit{CES Feedback}, \textit{Recommendations}, \textit{GSS Stats}, \textit{GSS Feedback}, \textit{Alumni Survey Stats} and \textit{Alumni Survey Feedback}. Each of these forms allows the management of data related to the feedback collected from current and graduating students, faculty members and alumni. The bulk of this data is utilized in indirect assessment and continuous program improvement. The forms provide the following functions:
\begin{itemize}
	\item \textit{CES Stats}: manage survey data gathered from course exit surveys (CESs) 
	\item \textit{CES Feedback}: manage count of CES participants and verbal student feedback 
	\item \textit{Recommendations}: manage feedback data gathered from faculty members 
	\item \textit{GSS Stats}: manage survey data gathered from graduating student surveys (GSSs)
	\item \textit{GSS Feedback}: manage count of GSS participants and verbal student feedback
	\item \textit{Alumni Survey Stats}: manage survey data gathered from alumni surveys
	\item \textit{Alumni Survey Feedback}: manage count of alumni survey participants and verbal alumni feedback
\end{itemize} 
Figure \ref{fig:recommend1} displays a data-entry sample using the \textit{Recommendations} form. It can be used to enter user feedback and link documentary evidence that can be used utilized for continuous improvement, as demonstrated in a Section \ref{sec:examples}.

\begin{remark}
	The forms described above are linked to individual tables having distinct relationship with other tables. These relationships are defined using Access GUI and fulfill  referential integrity constraints. For further details on the types of relationships and examples pertaining to ABET assessment in Microsoft Access, see \citep{Cliver2011}.
\end{remark}
\begin{remark}
	Form controls like text, list and combo boxes as well as buttons can be also inserted using Access's GUI. These controls can be assigned specific functions, like adding new records or displaying existing records, through macros - tasks performed routinely and repeatedly \citep{accessmacro_2018}.  
\end{remark}
\subsection{Reports}
Reports are the crux of ACT. They allow the database users and program evaluators to view the assessment data in an organized fashion and extract meaningful information from it. Several forms for generating customized reports are available under \textit{Reports-1} and \textit{Reports-2} tabs.
\subsubsection{Reports-1}
\begin{figure}[t!]
	\centering
	\includegraphics[width=\linewidth,trim=0cm 2cm 12cm 4.25cm,clip]{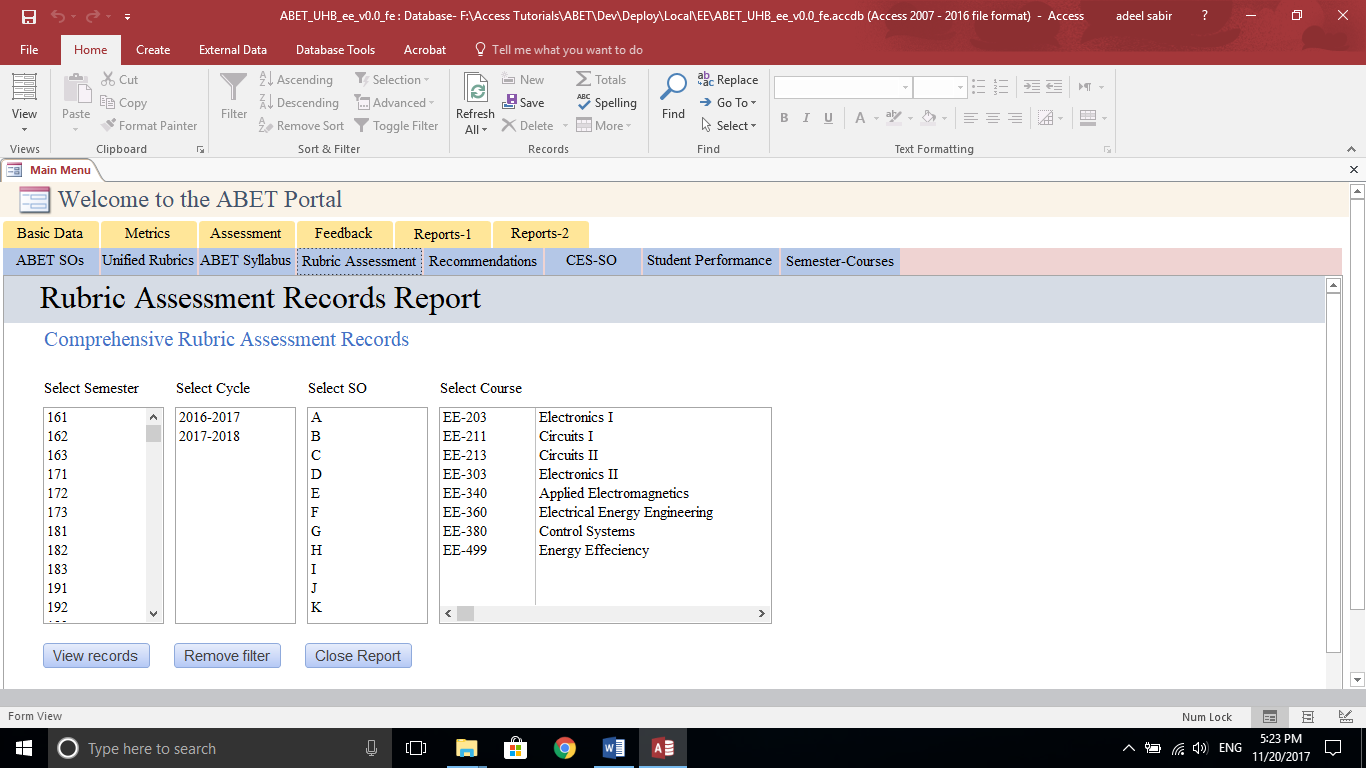}
	\caption{\textit{Rubric Assessment} report generation form.}
	\label{fig:rubricreportform_snap1}
\end{figure}

\begin{figure}[t!]
	\centering
	\includegraphics[width=\linewidth,trim=13cm 2.25cm 13cm 5cm,clip]{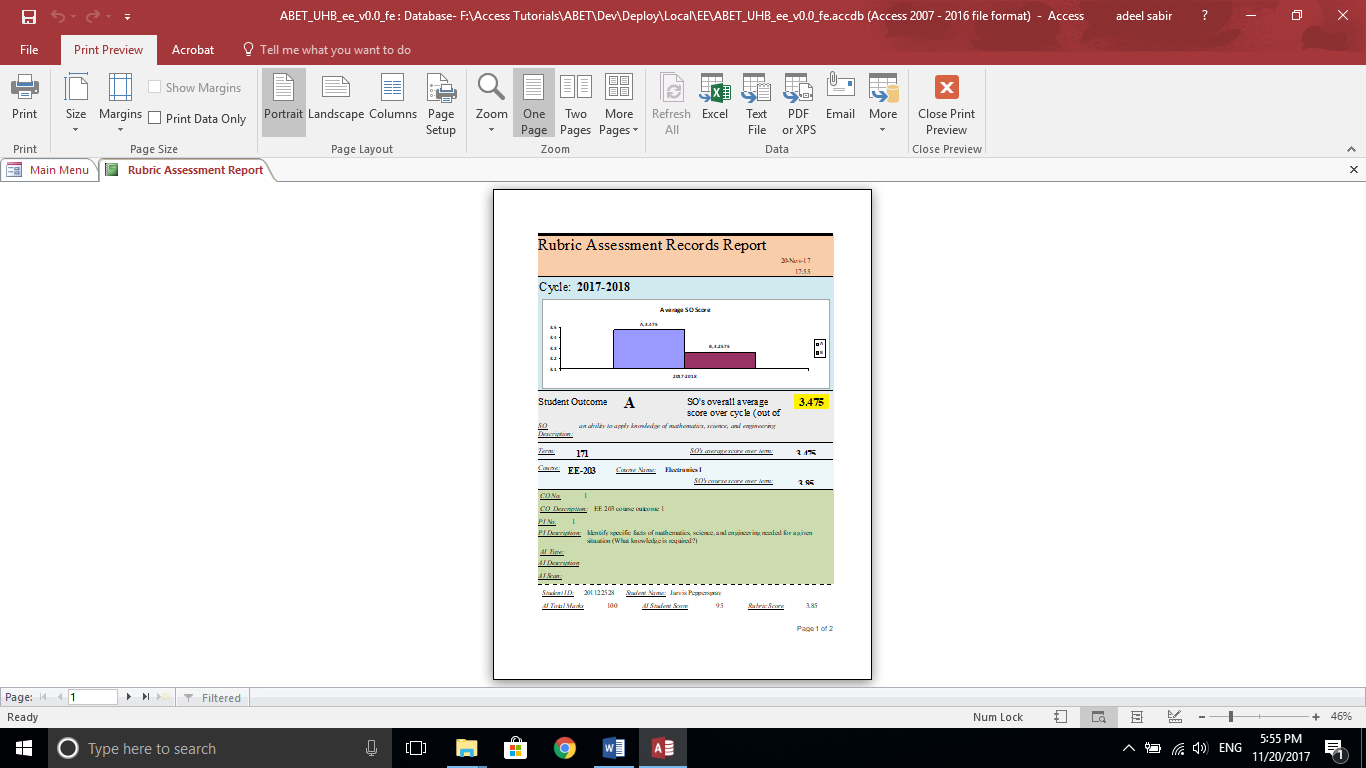}
	\caption{First page of a filtered report generated using the \textit{Rubric Assessment} form.}
	\label{fig:reportview_snap1}
\end{figure}

\begin{figure}[t!]
	\centering
	\includegraphics[width=\linewidth,trim=0cm 2cm 12cm 4.25cm,clip]{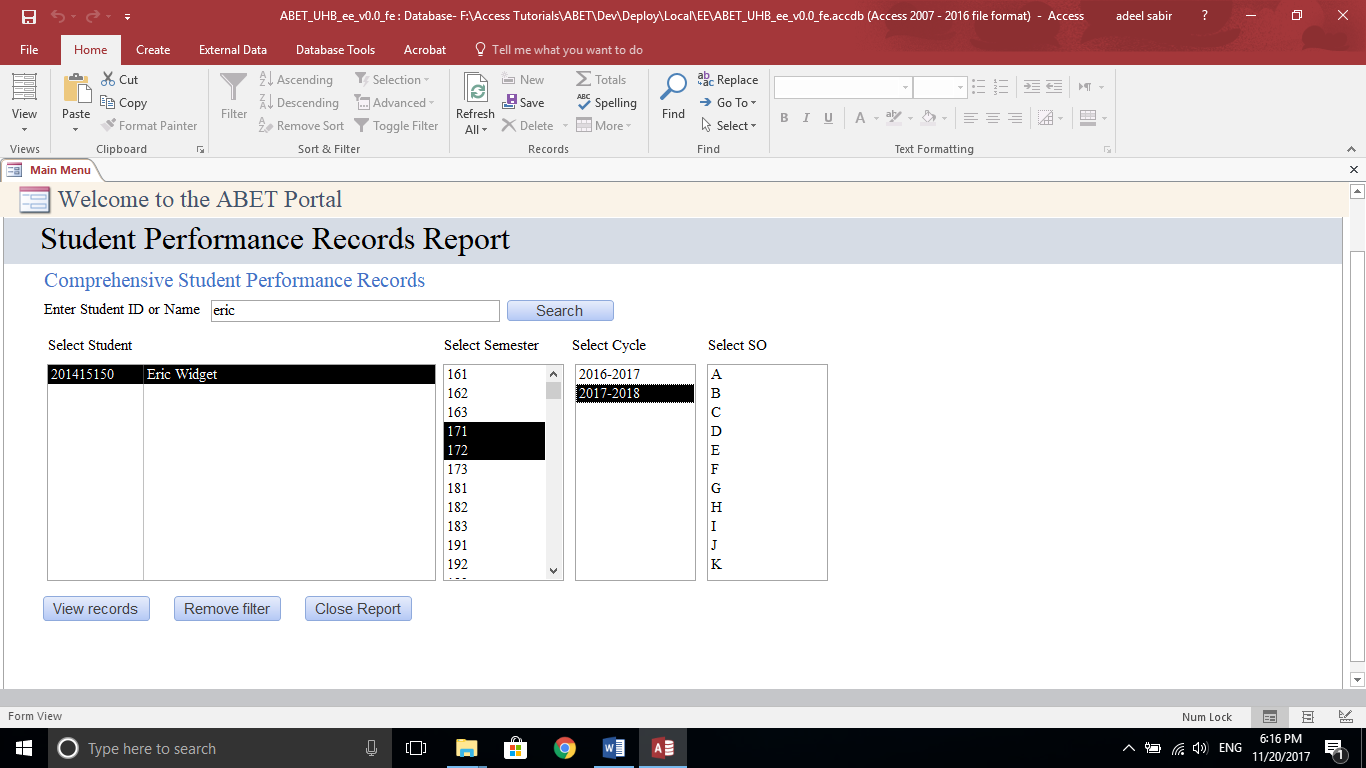}
	\caption{\textit{Student Performance} form showing the filtering criteria for report generation.}
	\label{fig:studentreportform_snap1}
\end{figure}

The \textit{Reports-1} tab houses eight reporting forms namely \textit{ABET SO}, \textit{Unified Rubrics}, \textit{ABET Syllabus}, \textit{Rubric Assessment}, \textit{CES-SO}, \textit{Student Performance} and \textit{Semester-Courses}. They provide the following functions:
\begin{itemize}
	\item \textit{ABET SO}: view student outcomes report
	\item \textit{Unified Rubrics}: view report of assessment rubic definitions and descriptions
	\item \textit{ABET Syllabus}: view course syllabuses and CO-SO mappings report
	\item \textit{Rubric Assessment}: view rubric assessment report
	\item \textit{Recommendation}: view faculty feedback report
	\item \textit{CES-SO}: view report of CESs on COs mapped to SOs
	\item \textit{Student Performance}: view a report of student performance on SOs
	\item \textit{Semester-Courses}: view report of courses offered by semesters and assessment cycles
\end{itemize} 
Each of these forms contains a combination of list, combo and text boxes allowing users to filter and customize the reports according to their needs. Figure \ref{fig:rubricreportform_snap1} shows the \textit{Rubric Assessment} report generation form with interactive list boxes for report customization. Figure \ref{fig:reportview_snap1} shows a customized assessment report showing the scores of SOs A and E over a specific term and the averaged scores over the assessment cycle. Figure \ref{fig:studentreportform_snap1} shows the \textit{Student Performance} with filtered list boxes for a hypothetical student, allowing the user to view this student's performance in specific (or all) SOs. 

\subsubsection{Reports-2}
\begin{figure}[t!]
	\centering
	\includegraphics[width=\linewidth,trim=1.75cm 2cm 15cm 4.6cm,clip]{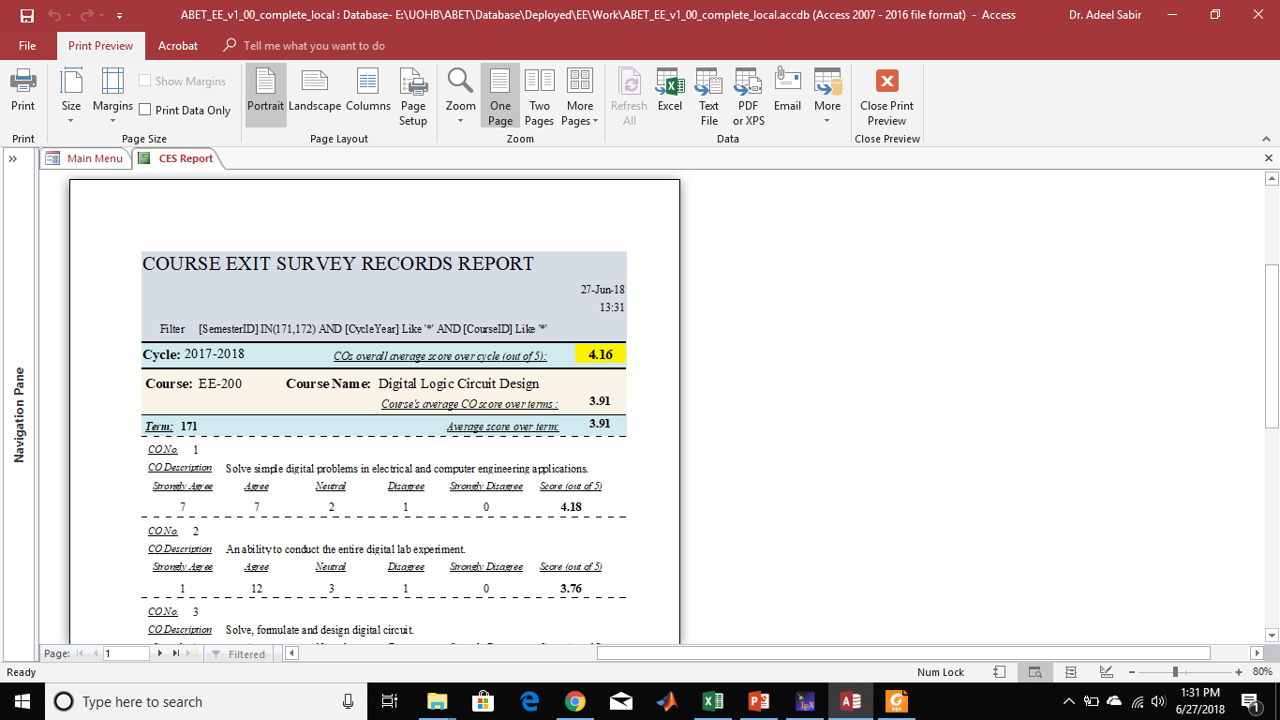}
	\caption{First page of a filtered report generated using the \textit{CES Stats} form.}
	\label{fig:cesreportview_snap1}
\end{figure}

The \textit{Reports-2} tab houses six forms namely namely \textit{CES Stats}, \textit{CES Feedback}, \textit{GSS Stats}, \textit{GSS Feedback}, \textit{Alumni Survey Stats} and \textit{Alumni Survey Feedback}. They generate reports for the data entered using the forms described in subsection \ref{ssec:feedback}. Their descriptions are as follows:
\begin{itemize}
	\item \textit{CES Stats}: view report of CES statistics
	\item \textit{CES Feedback}: view report of CES feedback
	\item \textit{GSS Stats}: view report of GSS statistics
	\item \textit{GSS Feedback}: view report of GSS feedback
	\item \textit{Alumni Survey Stats}: view report of alumni survey statistics
	\item \textit{Alumni Survey Feedback}: view report of alumni survey feedback
\end{itemize} 
Figure \ref{fig:cesreportview_snap1} depicts part of a report generated using the \textit{CES Stats} form, showing some numerical CO scores from indirect assessment. Similar to the direct assessment, the indirect scores are automatically calculated by the tool using a mathematical mapping rule.

\section{Act usage examples and results} \label{sec:examples}
In this section, the results obtained from application usage are demonstrated through practical examples from the EE department at the authors' institution.
\subsection{Example 1 - criterion 4} \label{ssec:example1}
In the CAEP document \citep{abet_eac2017}, criterion 4, continuous improvement, states that ``\textit{The program must regularly use appropriate, documented processes for assessing and evaluating the extent to which the student outcomes are being attained. The results of these evaluations must be systematically utilized as input for the continuous improvement of the program. Other available information may also be used to assist in the continuous improvement of the program.}'' As noted in \citep{Cliver2011}, ABET evaluators have commonly observed that many programs attempt to fulfill criterion 4 through a set of spontaneous activities near the accreditation visit rather than a regular and ongoing process throughout the assessment cycle; a practice that is counter to the key idea of `continuous improvement'. These observations definitively suggest the need for a uniform and systematic method whereby the areas needing attention or improvement are constantly  identified, steps are regularly taken to resolve the relevant issues and the underlying activities are documented. The following examples demonstrate how some feedback-driven improvements in the program were documented using the application. 

\subsubsection{Laboratory upgrade}
Student feedback for the course EE-340 (Applied Electromagnetics) prior to the term 171 suggested that the lab manual was not well-written and that most of the experiments were simulation based and not hardware based. This prompted a thorough investigation an as a result, the laboratory manual and available experimental equipment was inspected by two different faculty members. The following was observations were recorded using the \textit{Recommendations} feature:
\begin{itemize}
	\item[O1.] A number of the experiments in the lab manual were not properly written, had inadequate instructions and inconsistencies.
	\item[O2.] There was some unused hardware in the lab that was procured under an older purchase requisition (PR) overseen by a professor no longer with the department. The lab instructors were not familiar with its usage but there were experiments in the manual based on that hardware.
	\item[O3.] Some of the hardware from the PR was still not delivered.
	\item[O4.] There were insufficient number of computers in the laboratory as per the lab requirement.
\end{itemize}
Having identified these issues, the following actions were taken and documented in the database:
\begin{itemize}
	\item[A1.] The whole lab manual was revised and rewritten. Some older experiments were replaced with new ones and the discrepancies and inconsistencies were removed. It was ensured that students had clear and step-by-step instructions on what to do and what were the expected outcomes of the experiment.
	\item[A2.] The local supplier of the unused hardware was contacted and asked to provide training and technical support. They were also notified of the missing items yet to be delivered. 
	\item[A3.] The supplier sent in an engineer who set up the experimental equipment and provided training seminars to the faculty. They also delivered the missing items.
	\item[A4.] Following the hardware training seminars, new experiments were designed and included in the updated lab manual by the faculty members overseeing the issue.
	\item[A5.] The information technology (IT) department was contacted to provide additional computers for the lab on an urgent basis. The request was processed within a few days. 
\end{itemize}
The bulk of these corrective tasks were undertaken over the course of the summer 2017 and 171 terms, with some training sessions held in the following term 172. This exercise was well-received by the faculty as they availed an opportunity to get hands-on training on some interesting hardware and software tools. Their feedback on the revised lab manual was also positive. The renovation of the laboratory course resulted in no complains being received from the students regarding comprehension or the lab facility, as observed from student feedback in the course's offering in the succeeding term.

\begin{figure}[t!]
	\centering
	\includegraphics[width=\linewidth,trim=11.75cm 1.75cm 9cm 5cm,clip]{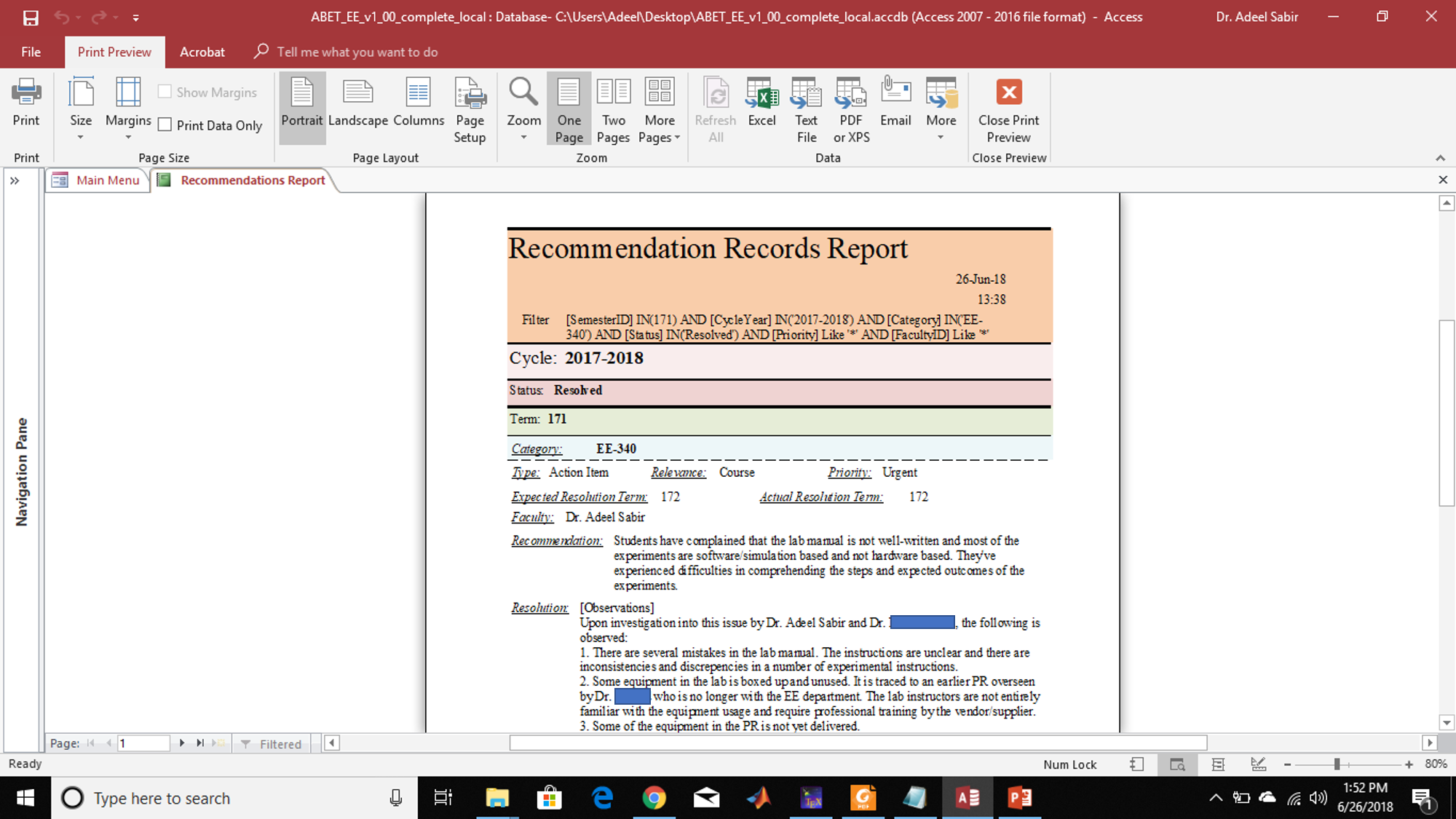}
	\caption{Recommendation report showing the documentation of a closed issue.}
	\label{fig:recommendreport_snap1}
\end{figure}

The entire process was periodically documented using the \textit{Recommendations} form by the appropriate faculty. Figure \ref{fig:recommendreport_snap1} shows part of the recommendations report documenting the process. It highlights various aspects of this exercise e.g., the chronological order of terms, relevance, focal person, type of actions required and priority level of the issue. Evidentiary material was also linked to this activity using the \textit{Recommendations} form under the \textit{Feedback} tab; an interactive link appears at the bottom of the report. The documentation generated as a result of these activities, along with the recommendations report, serve as evidence for continuous improvement.   

\subsection{Example 2 - criterion 3}
Criterion 3 relates directly to the students. It necessitates a documented method for assessing whether the they are adequately fulfilling certain academic goals formulated in the program's educational objectives and are satisfactorily attaining a specific set of outcomes listed in the CAEP document. ACT not only allows for quantitatively assessing ABET SOs through the method described in subsection \ref{ssec:characterization}, but also enables an assessor to draw parallels between direct and indirect data as well as examine the correlation between qualitative and quantitative data. The following examples demonstrate these abilities.

\subsubsection{Quantitative assessment and achievement of SOs}
\begin{figure}[t!]
	\centering
	\includegraphics[width=\linewidth]{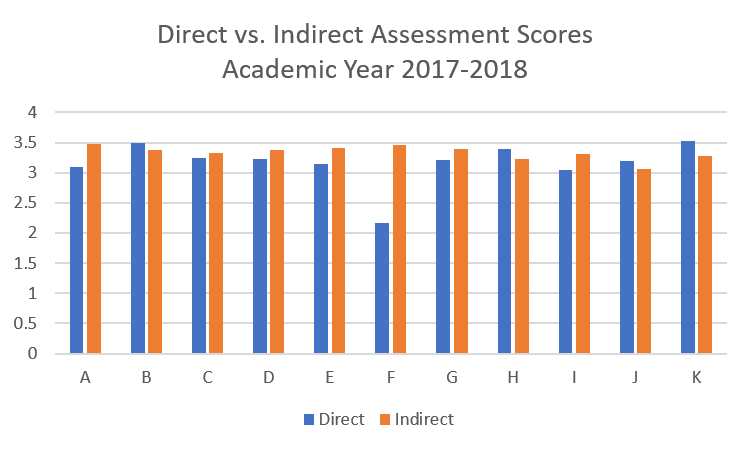}
	\caption{Student outcome achievement scores over academic year 2017-2018.}
	\label{fig:All_SOs_chart}
\end{figure}

Student outcomes were quantitatively assessed in ACT over two terms in the academic year 2017-2018. Figure \ref{fig:All_SOs_chart} shows the achievement scores for all the SOs, averaged over the academic year. The chart in the figure indicates that all outcomes met the achievement threshold of 2.4, except for SO F. Additionally, with the exception of SO F, all direct and indirect scores were close to each other, indicating a close correlation between student and instructor evaluations. 

Some key reasons were identified for the low score and lack of correlation in outcome F, which aims at developing \textit{an understanding of professional and ethical responsibility}. Primarily, the outcome was assessed only once, and through a single course that is prerequisite to the capstone design project. Students were exposed to ethics and professionalism in the engineering vocation merely under a theoretical framework - they were not able to adequately appreciate the dynamics of making engineering decisions and their impact on society. Therefore, despite the fact that the students perceived themselves to have satisfactory abilities to act responsibly and professionally, instructor evaluation indicated inadequate achievement of this SO; hence the large gap between direct and indirect scores. Insufficient number of courses covering this outcome was another reason behind its poor achievement.

In order to improve the achievement level in this SO, it was proposed that this outcome be covered in the capstone project course, whereby pupils came into contact with various ethical dilemmas while developing engineering solutions to real-life problems, and were more likely to develop improved insights into moral decision-making. Additionally, this SO was made part of the mandatory industrial cooperative training program so that the students had better opportunities to understand and appreciate the importance of professionalism and work ethics in the industry. Results of future evaluations will be analyzed to determine if achievement levels improve in response to these changes.

\subsubsection{Direct vs. indirect SO assessment correlation}
\begin{figure}[t!]
	\centering
	\includegraphics[width=\linewidth]{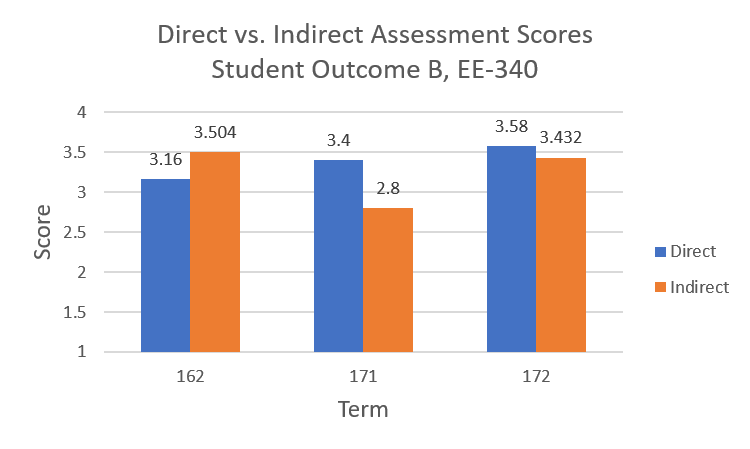}
	\caption{Summary of SO B scores for EE-340 for three consecutive terms.}
	\label{fig:SO_B_chart1}
\end{figure}

\begin{figure}[t!]
	\centering
	\includegraphics[width=\linewidth]{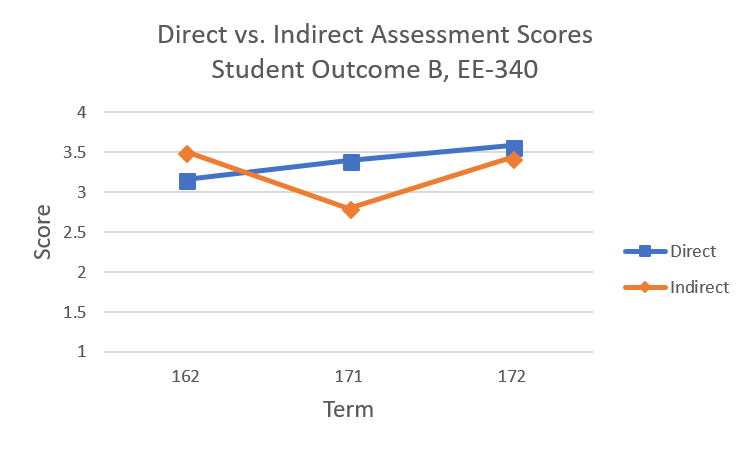}
	\caption{Trend of SO B scores for EE-340 over three consecutive terms.}
	\label{fig:SO_B_chart2}
\end{figure}

Student outcome B in a CAEP document mandates that the students must have ``\textit{an ability to design and conduct experiments, as well as to analyze and interpret data}''. Assessment information collected over three terms for SO B in the course EE-340 was compared and analyzed to determine whether a correlation could be found between direct and indirect data and/or, if the quantitative scores of the SO could be linked to the qualitative measures taken to improve the laboratory part of the course, described in the subsection \ref{ssec:example1}. Some statistics of the assessment data are depicted in Figures \ref{fig:SO_B_chart1} and \ref{fig:SO_B_chart2}. Indirect assessment scores are mapped to a scale of 1-4 for comparison with direct scores.

From Figure \ref{fig:SO_B_chart1}, the direct and indirect scores for SO B from terms 162 and 172 appear to be more closely correlated than the term 171. The chart shown in Figure \ref{fig:SO_B_chart2} indicates a gradual by steady improvement in the direct scores while the indirect scores exhibit a mixed behavior. The quantitative improvement in the direct assessment scores can, to an extent, be directly correlated to the qualitative measures taken to improve the laboratory part of EE-340 over this period. This inference is supported by an absence of negative feedback about the laboratory sessions and the manual, which was regularly observed prior to the implementation of a corrective action plan, described in subsection \ref{ssec:example1}. 

\begin{figure}[t!]
	\centering
	\includegraphics[width=\linewidth]{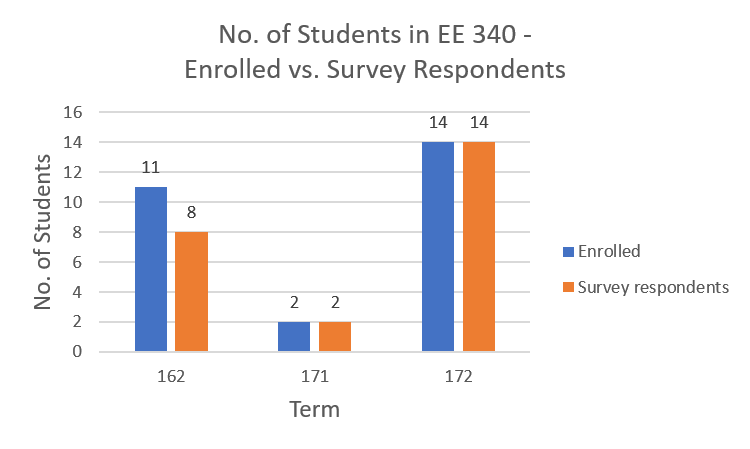}
	\caption{Enrollment and survey statistics of EE-340 over three consecutive terms.}
	\label{fig:SO_B_chart3}
\end{figure}

The recession in the indirect trend from 162 to 171 could be linked to the fact that the course was offered as a special circumstance for a graduating student and had an enrollment of only two students. It is normally offered only in the second term of an academic year. The enrollment and survey statistics for EE-340 are shown in Figure \ref{fig:SO_B_chart3}. The chart compares the number of students that completed the course versus those that responded to the exit survey.
It is regularly reported by the faculty that indirect assessment scores tend to exhibit anomalous behavior for courses having less than average enrollment or a low count on student survey responses. Based on past experience and departmental trends, it has been observed that a larger number of responses on student surveys helps understand student perception more accurately than when the response-count is low. When participation in student surveys is scarce, anomalies may arise due to student assessments lying on a narrow perception spectrum. The chart in Figure \ref{fig:SO_B_chart3} indicates a sharp contrast between the enrollment and survey statistics between consecutive terms. This contrast was also reflected in the indirect trends of the SO, depicted in Figure \ref{fig:SO_B_chart2}. On the other hand, terms 162 and 171 had a higher enrollment and survey count than term 171 - their indirect scores were also comparable and higher than term 171.    

\section{Conclusion and future work} \label{sec:conclusion}
Fulfilling the requirements for ABET accreditation can be a daunting undertaking, and requires collaboration and team effort from all faculty members. Automation of the underlying tasks using electronic tools increases process efficiency and reduces paperwork. It also encourages broader faculty participation and enables the ABET accreditation committee to manage the assessment activity more effectively in comparison to a paper-based procedure. The use of electronic tools enables the faculty members and evaluators to extract meaningful information from large amounts of assessment data, and facilitates the identification and implementation of corrective actions and improvement plans. However, building a computerized assessment application is, in and of itself, a challenging project. This article demonstrates how the intensive accreditation activities can be made efficient and paperless by the application of ACT - an electronic database tool built using Microsoft Access.

The whole endeavor was not without its challenges. While ACT was well-received by most of the faculty members, some were still reluctant in fully utilizing the application, primarily due to lack of adequate assessment data, statistics and analytical results. This issue is currently being actively addressed. It is expected that gathering more data over the future terms, utilizing it for further analysis and exposition of quantifiable improvements linked to faculty contributions will increase their awareness about the usefulness of the assessment tool and motivate its usage. A statistical comparison between the paperwork and man-hours of the older, paper-based process and the new electronic process is also planned to validate efficiency gains. Furthermore, the authors plan to incorporate the program's educational objectives into the application and map them to ABET SOs to further enrich the quantitative analysis. Collection of additional indirect feedback (e.g., from employers, university's board of directors) and mapping it to student outcomes will also be made a part of future upgrades to the tool.

\bibliographystyle{agsm} % or use agsm style
%\bibliography{\jobname} % <======= to use bib file created with filecontents
\bibliography{library}

\section*{About the authors}
\textit{Adeel Sabir} is an Assistant Professor with the Department of Electrical Engineering, University of Hafr Al Batin. He received his Ph.D. from King Fahd University of Petroleum and Minerals (KFUPM) in 2016 with a specialization in power and control systems. He received the Exceptional Laboratory Coordinator award while serving as the coordinator for power system laboratory courses at KFUPM. He also received the Excellent Teaching Performance award three times at KFUPM ($2^{nd}$ term 2011-2012 \& 2013-2015, and $1^{st}$ term 2014-2015) while teaching the laboratory sessions for control engineering and power systems undergraduate courses. His research interests include application of technology and automation tools in engineering education, robust control of renewable energy systems, and smart grids. \newline\newline
\textit{Nisar A. Abbasi} was born in Abbottabad, Pakistan. He received the B.Eng. degree (with honors) in electrical and electronic engineering from the University of Engineering and Technology, Peshawar, Pakistan, in 2001, the M.Sc. and Ph.D. degrees in electronic engineering from the University of Sheffield, Sheffield, U.K., in 2007 and 2011 respectively. He has held various positions such as Assistant Manager, Manager and General Manager while working for The Ministry of Defense, Pakistan between 2001 and 2016. He was a visiting faculty at The University of Engineering and Technology, Taxila, Pakistan for five years between 2011 and 2016. He is currently an Assistant Professor at The University of Hafr Al Batin, Hafr Al Batin, KSA. His research interests cover the design and analysis of various antenna types. Dr. Abbasi is a Professional Engineer (PE) in Pakistan and a member of Pakistan Engineering Council (PEC).\newline\newline
\textit{Md Nurul Islam} received his Ph.D. from the University of Newcastle, Australia in 2013. He is Currently working as an Assistant Professor, Electrical Engineering at the University of Hafr Al-Batin. His research interests include linear and nonlinear control, and multi-link robotics and control.
\end{document}